\documentclass[prb,11t]{revtex4}
\usepackage{amsfonts}
\usepackage[T1]{fontenc}
\usepackage{amsmath,amsbsy,amssymb,graphicx}
\usepackage{times}

\begin{document}

\title{ Variational Quantum Support Vector Machine based on $\Gamma$ matrix
expansion\\
and Variational Universal-Quantum-State Generator}
\author{Motohiko Ezawa}
\affiliation{Department of Applied Physics, University of Tokyo, Hongo 7-3-1, 113-8656,
Japan}

\begin{abstract}
We analyze a binary classification problem by using a support vector machine
based on variational quantum-circuit model. We propose to solve a linear
equation of the support vector machine by using a $\Gamma$ matrix expansion.
In addition, it is shown that an arbitrary quantum state is prepared by
optimizing a universal quantum circuit representing an arbitrary $U(2^N)$
based on the steepest descent method. It may be a quantum generalization of
Field-Programmable-Gate Array (FPGA).
\end{abstract}

\maketitle

\medskip\noindent\textbf{\large Introduction}\medskip

Quantum computation is a hottest topic in contemporary physics\cite%
{Feynman,DiVi,Nielsen}. An efficient application of quantum computations is
machine learning, which is called quantum machine learning\cite%
{Lloyd,Schuld,Biamonte,Wittek,Harrow,Wiebe,Reben,ZLi,SchuldB,Hav,Lamata,Cong}. 
A support vector machine is one of the most fundamental algorithms for
machine learning\cite{Vap,Noble,Suy}, which classifies data into two classes
by a hyperplane. The optimal hyperplane is determined by an associated
linear equation $F|\psi _{\text{in}}\rangle =|\psi _{\text{out}}\rangle $,
where $F$ and $|\psi _{\text{out}}\rangle $ are given. A quantum support
vector machine solves this linear equation by a quantum computer\cite{Zhaokai,Hav,Reben}. 
Usually, the linear equation is solved by the
Harrow-Hassidim-Lloyd (HHL) algorithm\cite{HHL}. However, this algorithm
requires many quantum gates. Thus, the HHL algorithm is hard to be executed
by using a near-term quantum computer. Actually, this algorithm has
experimentally been verified only for two and three qubits\cite{XCai,Barz,JPan}. 
In addition, it requires a unitary operator to execute $e^{iFt}$, which is quite hard to be implemented.

The number of qubits in current quantum computers is restricted. Variational
quantum algorithms are appropriate for these small-qubit quantum computers,
which use both quantum computers and classical computers. Various methods
have been proposed such as Quantum Approximate Optimization Algorithm (QAOA)\cite{QAOA}, 
variational eigenvalue solver\cite{Peru}, quantum circuit
learning\cite{Mitarai} and quantum linear solver\cite{Prie,Xxu}. We use wave
functions with variational parameters in QAOA, which are optimized by
minimizing the expectation value of the Hamiltonian. A quantum circuit has
variational parameters in quantum circuit learning\cite{Mitarai}, which are
optimized by minimizing a certain cost function. A quantum linear solver
solves a linear equation by variational ansatz\cite{Prie,Xxu}. The simplest
method of the optimization is a steepest-descent method.

In this paper, we present a variational method for a quantum support vector
machine by solving an associated linear equation based on variational
quantum circuit learning. We propose a method to expand the matrix $F$ by
the $\Gamma $ matrices, which gives simple quantum circuits. We also propose
a variational method to construct an arbitrary state by using a universal
quantum circuit to represent an arbitrary unitary matrix $U(2^{N})$. We
prepare various internal parameters for a universal quantum circuit, which
we optimize by minimizing a certain cost function. Our circuit is capable to
determine the unitary transformation $U$ satisfying $U|\psi _{\text{initial}}\rangle =|\psi _{\text{final}}\rangle $ 
with arbitrary given states $|\psi_{\text{initial}}\rangle $ and $|\psi _{\text{final}}\rangle $. It will be a
quantum generalization of field-programmable-gate array (FPGA), which may
execute arbitrary outputs with arbitrary inputs.\medskip

\medskip\noindent \textbf{\large Risults}

\medskip \noindent \textbf{Support vector machine.}\smallskip

\begin{figure}[t]
\centerline{\includegraphics[width=0.88\textwidth]{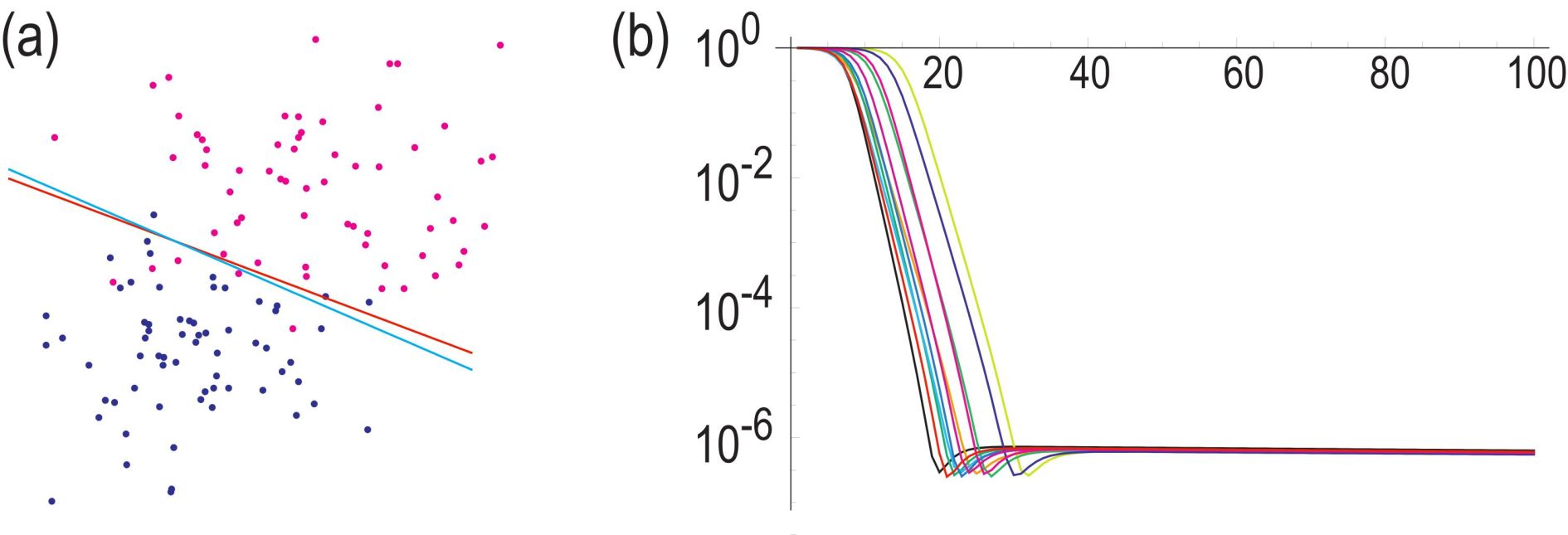}}
\caption{(a) Binary classification of red and blue points based on a quantum
support vector machine with soft margin. A magenta (cyan) line obtained
by an exact solution (variational method). (b) Evolution of the cost
function. The vertical axis is the log$_{10}E_{\text{cost}}$. The horizontal
axis is the variational step number. We have used $r=2$, $\protect\xi_{1}=0.001$ 
and $\protect\xi _{2}=0.0005$ and $\protect\gamma =1$. We have
runed simulations ten times. }
\label{FigSVM}
\end{figure}

A support vector machine (SVM) is a computer algorithm that learns by
examples to assign labels to objects. It is a typical method to solve a
binary-classification problem\cite{Vap}. A simplest example reads as
follows. Suppose that there are red and blue points whose distributions are
almost separated into two dimensions. We classify these data points into two
classes by a line, as illustrated in Fig.\ref{FigSVM}.

In general, $M$ data points are spattered in $D$ dimensions, which we denote 
$\boldsymbol{x}_{j}$, where $1\leq j\leq M$. The problem is to determine a
hyperplane,%
\begin{equation}
\boldsymbol{\omega }\cdot \boldsymbol{x}+\omega _{0}=0,
\end{equation}%
separating data into two classes with the use of a support vector machine.
We set 
\begin{equation}
\boldsymbol{\omega }\cdot \boldsymbol{x}+\omega _{0}>0
\end{equation}%
for red points and 
\begin{equation}
\boldsymbol{\omega }\cdot \boldsymbol{x}+\omega _{0}<0
\end{equation}%
for blue points. These conditions are implemented by introducing a function%
\begin{equation}
f\left( \boldsymbol{x}\right) =\text{sgn}\left( \boldsymbol{\omega }\cdot \boldsymbol{x}%
+\omega _{0}\right) ,
\end{equation}%
which assigns $f\left( \boldsymbol{x}\right) =1$ to red points and $f\left( 
\boldsymbol{x}\right) =-1$ to blue points. In order to determine $\omega _{0}$
and $\boldsymbol{\omega }$ for a given set of data $\boldsymbol{x}_{j}$, we
introduce real numbers $\alpha _{j}$ by%
\begin{equation}
\boldsymbol{\omega }=\sum_{j=1}^{M}\alpha _{j}\boldsymbol{x}_{j}.
\end{equation}%
A support vector machine enables us to determine $\omega _{0}$ and $\alpha
_{j}$ by solving the linear equation%
\begin{equation}
F\left( 
\begin{array}{c}
\omega _{0} \\ 
\alpha _{1} \\ 
\vdots \\ 
\alpha _{M}%
\end{array}%
\right) =\left( 
\begin{array}{c}
0 \\ 
y_{1} \\ 
\vdots \\ 
y_{M}%
\end{array}%
\right) ,  \label{LinearEq}
\end{equation}%
where $y_{i}=f(x_{i})=\pm 1$, and $F$ is a $(M+1)\times (M+1)$ matrix given
by%
\begin{equation}
F=\left( 
\begin{array}{cccc}
0 & 1 & \cdots & 1 \\ 
1 &  &  &  \\ 
\vdots &  & K+I_{M}/\gamma &  \\ 
1 &  &  & 
\end{array}%
\right) .
\end{equation}%
Here, 
\begin{equation}
K_{ij}=\boldsymbol{x}_{i}\cdot \boldsymbol{x}_{j},
\end{equation}%
is a Kernel matrix, and $\gamma $ is a certain fixed constant which assures
the existence of the solution of the linear equation (\ref{LinearEq}) even
when the red and blue points are slightly inseparable. Note that $\gamma
\rightarrow \infty $ corresponds to the hard margin condition. Details of
the derivation of Eq.(\ref{LinearEq}) are given in Method A.

\medskip\noindent\textbf{Quantum linear solver based on $\Gamma$ matrix expansion.}\smallskip

We solve the linear equation (\ref{LinearEq}) by a quantum computer. In
general, we solve a linear equation%
\begin{equation}
F\left\vert \psi _{\text{in}}\right\rangle =c\left\vert \psi _{\text{out}%
}\right\rangle ,  \label{EqA}
\end{equation}%
for an arbitrary given non-unitary matrix $F$ and an arbitrary given state $%
\left\vert \psi _{\text{out}}\right\rangle $. Here, the coefficient $c$ is
introduced to preserve the norm of the state, and it is given by 
\begin{equation}
c=\sqrt{\left\langle \psi _{\text{in}}\right\vert F^{\dagger }F\left\vert
\psi _{\text{in}}\right\rangle }.
\end{equation}%
The HHL algorithm\cite{HHL} is a most famous algorithm to solve this linear
equation by a quantum computer. We first construct a Hermitian matrix by%
\begin{equation}
H=\left( 
\begin{array}{cc}
0 & F \\ 
F^{\dagger } & 0%
\end{array}%
\right) .
\end{equation}%
Then, a unitary matrix associated with $F$ is uniquely obtained by $e^{iHt}$. 
Nevertheless, it requires many quantum gates. In addition, it is a
nontrivial problem to implement $e^{iHt}$.

Recently, variational methods have been proposed\cite{Prie} to solve the
linear equation (\ref{EqA}). In one of the methods, the matrix $F$ is
expanded in terms of some unitary matrices $U_{j}$ as%
\begin{equation}
F=\sum_{j=0}^{2^{N}-1}c_{j}U_{j}.
\end{equation}%
In general, a complicated quantum circuit is necessary to determine the
coefficient $c_{j}$.

We start with a trial state $|\tilde{\psi}_{\text{in}}\rangle $ to
determine the state $|\psi _{\text{in}}\rangle $. Application of each
unitary matrix to this state is efficiently done by a quantum computer, $%
U_{j}|\tilde{\psi}_{\text{in}}\rangle =|\tilde{\psi}_{\text{out}}^{\left(
j\right) }\rangle $, and we obtain%
\begin{equation}
F|\tilde{\psi}_{\text{in}}\rangle =\sum_{j=0}^{2^{N}-1}c_{j}U_{j}|\tilde{\psi%
}_{\text{in}}\rangle =\sum_{j=0}^{2^{N}-1}c_{j}|\tilde{\psi}_{\text{out}%
}^{\left( j\right) }\rangle \equiv c|\tilde{\psi}_{\text{out}}\rangle ,
\end{equation}%
where $|\tilde{\psi}_{\text{out}}\rangle $ is an approximation of the given
state $\left\vert \psi _{\text{out}}\right\rangle $. We tune a trial state 
$|\tilde{\psi}_{\text{in}}\rangle $ by a variational method so as to
minimize the cost function\cite{Prie}%
\begin{equation}
E_{\text{cost}}\equiv 1-\left\vert \langle \tilde{\psi}_{\text{out}}|\psi _{%
\text{out}}\rangle \right\vert ^{2},
\end{equation}%
which measures the similarity between the approximate state $|\tilde{\psi}_{\text{out}}\rangle $ 
and the state $\left\vert \psi _{\text{out}}\right\rangle $ in (\ref{EqA}). We have $0\leq E_{\text{cost}}\leq 1$,
where $E_{\text{cost}}=0$ for the exact solution. The merit of this cost
function is that the inner product is naturally calculated by a quantum
computer.

Let the dimension of the matrix $F$ be $2^{N}$. It is enough to use $N$
satisfying $2^{N-1}<D\leq 2^{N}$ without loss of generality by adding
trivial $2^{N}-D$ components to the linear equation. We propose to expand
the matrix $F$ by the gamma matrices $\Gamma _{j}$ as%
\begin{equation}
F=\sum_{j=0}^{2^{N}-1}c_{j}\Gamma _{j},
\end{equation}%
with 
\begin{equation}
\Gamma _{j}=\bigotimes_{\beta =1}^{N}\sigma _{\alpha }^{\left( \beta \right)
},
\end{equation}%
where $\alpha $ $=0,x,y$ and $z$.

The merit of our method is that it is straightforward to determine $c_{j}$
by the well-known formula%
\begin{equation}
c_{j}=\text{Tr}\left[ \Gamma _{j}F\right] .  \label{cj}
\end{equation}%
In order to construct a quantum circuit to calculate $c_{j}$, we express the
matrix $F$ by column vectors as%
\begin{equation}
F=\left\{ \left\vert f_{0}\right\rangle ,\cdots ,\left\vert
f_{2^{N}-1}\right\rangle \right\} .
\end{equation}%
We have $\left( \left\vert f_{q-1}\right\rangle \right) _{p}=F_{pq}$, where
subscript $p$ denotes the $p$-th component of $\left\vert
f_{q-1}\right\rangle $. Then $c_{j}$ is given by%
\begin{equation}
c_{j}=\sum_{q=0}^{2^{N}-1}\left( \Gamma _{j}\left\vert f_{q}\right\rangle
\right) _{q}=\sum_{q=0}^{2^{N}-1}\left\langle \!\left\langle q\right\vert
\right. \Gamma _{j}\left\vert f_{q}\right\rangle ,
\end{equation}%
where the subscript $q$ denotes the ($q+1$)-th component of $\Gamma
_{j}\left\vert f_{q}\right\rangle $. We have introduced a notation 
$\left\vert q\right\rangle \!\rangle \equiv |n_{1}n_{2}\cdots n_{N}\rangle $
with $n_{i}=0,1$, where $q$ is the decimal representation of the binary
number $n_{1}n_{2}\cdots n_{N}$. See explicit examples for one and two
qubits in Method B.

\begin{figure}[t]
\centerline{\includegraphics[width=0.88\textwidth]{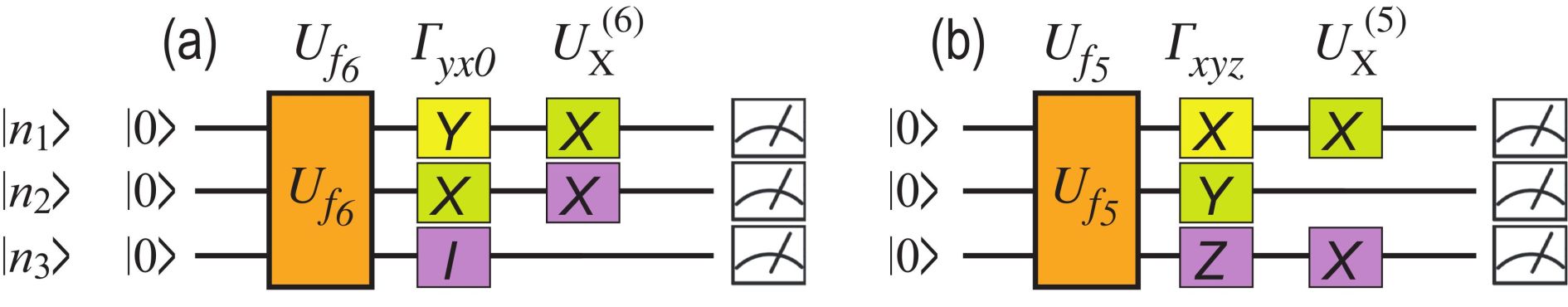}}
\caption{Quantum circuits determining $c_{j}$. We show an example with (a) 
$\Gamma _{yx0}=\protect\sigma _{y}\otimes \protect\sigma _{x}\otimes \protect\sigma _{0}$. 
$U_{X}^{\left( 6\right) }\left. \left\vert 0\right\rangle
\!\right\rangle =\protect\sigma _{x}^{\left( 1\right) }\protect\sigma_{x}^{\left( 2\right) }\left\vert 000\right\rangle =\left\vert
110\right\rangle =\left. \left\vert 6\right\rangle \!\right\rangle $ and (b) 
$\Gamma _{xyz}=\protect\sigma _{x}\otimes \protect\sigma _{y}\otimes \protect\sigma _{z}$. 
$U_{X}^{\left( 5\right) }\left\vert 0\right\rangle \!\rangle 
=\protect\sigma _{x}^{\left( 1\right) }\protect\sigma _{x}^{\left( 3\right)
}\left\vert 000\right\rangle =\left\vert 101\right\rangle =\left. \left\vert
5\right\rangle \!\right\rangle $.}
\label{FigUGamma}
\end{figure}

The state $\left. \left\vert q\right\rangle \!\right\rangle \equiv
|n_{1}n_{2}\cdots n_{N}\rangle $ is generated as follows. We prepare the NOT
gates $\sigma _{x}^{\left( i\right) }$ for the $i$-th qubit if $n_{i}=1$.
Using all these NOT gates we define%
\begin{equation}
U_{X}^{\left( q\right) }=\bigotimes\limits_{n_{i}=1}\sigma _{x}^{\left(
i\right) }.
\end{equation}%
We act it on the initial state $\left\vert 0\right\rangle \!\rangle $ and
obtain%
\begin{equation}
U_{X}^{\left( q\right) }\left\vert 0\right\rangle \!\rangle =\left\vert
q\right\rangle \!\rangle .
\end{equation}%
Next, we construct a unitary gate $U_{f_{q}}$\ generating $\left\vert
f_{q}\right\rangle $,%
\begin{equation}
U_{f_{q}}\left\vert 0\right\rangle \!\rangle =\left\vert f_{q}\right\rangle .
\end{equation}%
We will discuss how to prepare $U_{f_{q}}$ by a quantum circuit soon later;
See Eq.(\ref{EqB}). By using these operators, $c_{j}$ is expressed as%
\begin{equation}
c_{j}=\sum_{q=0}^{2^{N}-1}\left\langle \!\left\langle 0\right\vert \right.
U_{X}^{\left( q\right) }\Gamma _{j}U_{f_{q}}\left\vert 0\right\rangle
\!\rangle ,
\end{equation}%
which can be executed by a quantum computer. We show explicit examples in
Fig.\ref{FigUGamma}.

Once we have $c_{j}$, the final state is obtained by applying $\Gamma _{j}$
to $|\tilde{\psi}_{\text{in}}\rangle $ and taking sum over $j$, which leads
to%
\begin{equation}
|\tilde{\psi}_{\text{out}}\rangle =F|\tilde{\psi}_{\text{in}}\rangle
=\sum_{j=0}^{2^{N}-1}c_{j}\Gamma _{j}|\tilde{\psi}_{\text{in}}\rangle .
\end{equation}%
The implementation of the $\Gamma $ matrix is straightforward in quantum
circuit, because the $\Gamma $ matrix is composed of the Pauli sigma
matrices, as shown in Fig.\ref{FigUGamma}.

We may use the steepest descent method to find an optimal trial state 
$|\tilde{\psi}_{\text{in}}\rangle $ closest to the state $|\psi _{\text{in}}\rangle $. 
We calculate the difference of the cost function $\Delta E_{\text{cost}}$ 
when we slightly change the trial state $|\tilde{\psi}_{\text{in}}(t)\rangle $ 
at step $t$ by the amount of $\Delta |\tilde{\psi}_{\text{in}}(t)\rangle $ as%
\begin{equation}
\Delta E_{\text{cost}}\equiv E_{\text{cost}}\left( |\tilde{\psi}_{\text{in}%
}(t)\rangle +\Delta |\tilde{\psi}_{\text{in}}(t)\rangle \right) -E_{\text{%
cost}}\left( |\tilde{\psi}_{\text{in}}(t)\rangle \right) \simeq \frac{\Delta
E_{\text{cost}}}{\Delta |\tilde{\psi}_{\text{in}}(t)\rangle }\Delta |\tilde{%
\psi}_{\text{in}}(t)\rangle .
\end{equation}%
We explain how to construct $|\tilde{\psi}_{\text{in}}(t)\rangle $ by a
quantum circuit soon later; See Eq.(\ref{EqB}). Then, we renew the state as%
\begin{equation}
|\tilde{\psi}_{\text{in}}(t)\rangle \rightarrow |\tilde{\psi}_{\text{in}%
}(t)\rangle -\eta _{t}\frac{\Delta E_{\text{cost}}}{\Delta |\tilde{\psi}_{%
\text{in}}(t)\rangle }\Delta |\tilde{\psi}_{\text{in}}(t)\rangle ,
\end{equation}%
where we use an exponential function for $\eta _{t}$,%
\begin{equation}
\eta _{t}=\xi _{1}e^{-\xi _{2}t}.
\end{equation}%
We choose appropriate constants $\xi _{1}$\ and $\xi _{2}$\ for an efficient
search of the optimal solution, whose explicit examples are given in the
caption of Fig.\ref{FigUGamma}. We stop the renewal of the variational step
when the difference $\Delta |\tilde{\psi}_{\text{in}}(t)\rangle $ becomes
sufficiently small, which gives the optimal state of the linear equation (\ref{EqA}).

\medskip\noindent\textbf{Variational universal-quantum-state generator.}\smallskip

In order to construct the trial state $|\tilde{\psi}_{\text{in}}(t)\rangle $, 
it is necessary to prepare an arbitrary state $\left\vert \psi
\right\rangle $ by a quantum circuit. Alternatively, we need such a unitary
transformation $U$ that%
\begin{equation}
U\left\vert 0\right\rangle \!\rangle =\left\vert \psi \right\rangle .
\label{EqB}
\end{equation}%
It is known that any unitary transformation is done by a sequential
application of the Hadamard, the $\pi /4$ phase-shift and the CNOT gates\cite{Deutsch,Dawson}. 
Indeed, an arbitrary unitary matrix is decomposable into a
sequential application of quantum gates\cite{Deutsch,Dawson}, each of which
is constructed as a universal quantum circuit systematically\cite{Kraus,Vidal,Motto,Shende,Vatan,Sousa}. 
Universal quantum circuits have so
far been demonstrated experimentally for two and three qubits\cite{Hanne,DiCarlo,Qiang,Roy}.

We may use a variational method to construct $U$ satisfying Eq.(\ref{EqB}).
Quantum circuit learning is a variational method\cite{Mitarai}, where angle
variables $\theta _{i}$ are used as variational parameters in a quantum
circuit $U$, and the cost function is optimized by tuning $\theta _{i}$. We
propose to use a quantum circuit learning for a universal quantum circuit.
We show that an arbitrary state $\left\vert \psi \left( \theta _{i}\right)
\right\rangle $ can be generated by tuning $U\left( \theta _{i}\right) $
starting from the initial state $\left\vert 0\right\rangle \!\rangle $ as%
\begin{equation}
U\left( \theta _{i}\right) \left\vert 0\right\rangle \!\rangle =\left\vert
\psi \left( \theta _{i}\right) \right\rangle .
\end{equation}%
We adjust $\theta _{i}$ by minimizing the cost function%
\begin{equation}
E_{\text{cost}}\left( \theta _{i}\right) \equiv 1-\left\vert \left\langle
\psi \left( \theta _{i}\right) \left\vert \psi \right\rangle \right.
\right\vert ^{2},
\end{equation}%
which is the same as that of the variational quantum support vector machine.
We present explicit examples of universal quantum circuits for one, two and
three qubits in Method C.

\begin{figure}[t]
\centerline{\includegraphics[width=0.88\textwidth]{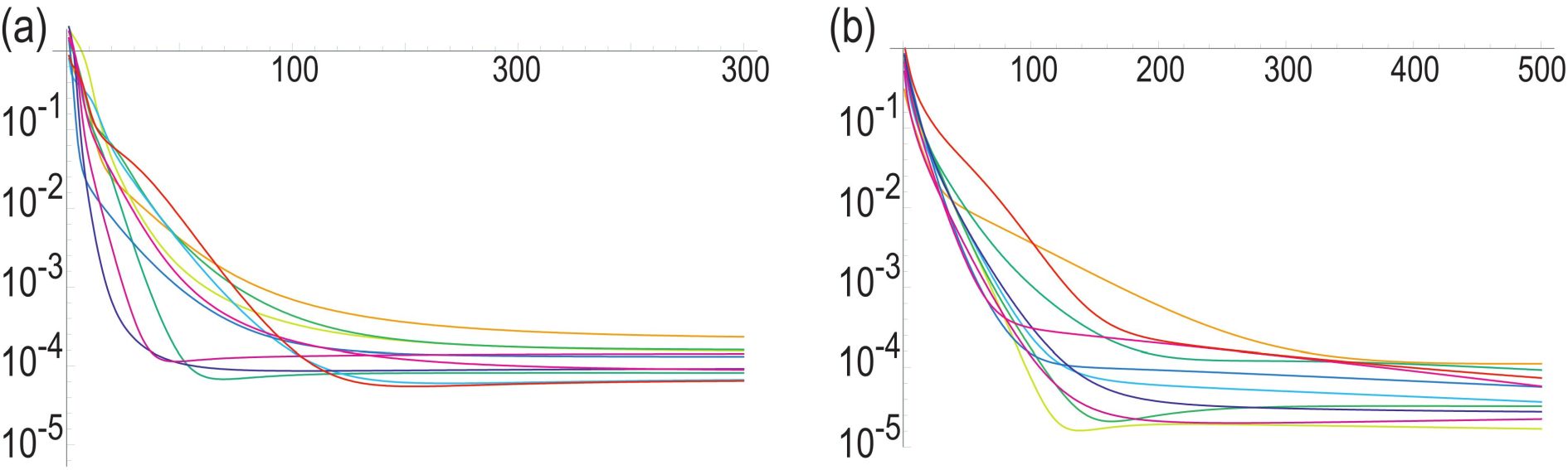}}
\caption{Evolution of the cost function for (a) two qubits and (b) three
qubits. The vertical axis is the log$_{10}E_{\text{cost}}$. The horizontal
axis is the number of variational steps. We use $c_{1}=0.005$ and 
$c_{2}=0.005$ for both the two- and three-qubit universal quantum circuits.
We prepare random initial and final states, where we have runed simulations
ten times. }
\label{FigCost}
\end{figure}

\medskip \noindent \textbf{Quantum Field-Programmable-Gate Array.}\smallskip

We next consider a problem to find a unitary transformation $U_{\text{ini-fin}}$ 
which maps an arbitrary initial state $\left\vert \psi _{\text{initial}}\right\rangle $ 
to an arbitrary final state $\left\vert \psi _{\text{final}}\right\rangle $,
\begin{equation}
U_{\text{ini-fin}}\left\vert \psi _{\text{initial}}\right\rangle =\left\vert
\psi _{\text{final}}\right\rangle .  \label{U1}
\end{equation}%
Since we can generate an arbitrary unitary matrix as in Eq.(\ref{EqB}), it
is possible to generate such matrices $U_{\text{ini}}$ and $U_{\text{fin}}$ that%
\begin{equation}
U_{\text{ini}}\left\vert 0\right\rangle \!\rangle =\left\vert \psi _{\text{%
initial}}\right\rangle ,\qquad U_{\text{fin}}\left\vert 0\right\rangle
\!\rangle =\left\vert \psi _{\text{final}}\right\rangle .  \label{U3}
\end{equation}%
Then, Eq.(\ref{U1}) is solved as 
\begin{equation}
U_{\text{fin}}=U_{\text{ini-fin}}U_{\text{ini}},
\end{equation}%
since $U_{\text{ini-fin}}\left\vert \psi _{\text{initial}}\right\rangle 
=U_{\text{ini-fin}}U_{\text{ini}}\left\vert 0\right\rangle \!\rangle =\left\vert
\psi _{\text{final}}\right\rangle =U_{\text{fin}}\left\vert 0\right\rangle
\!\rangle $.

An FPGA is a classical integrated circuit, which can be programmable by a
customer or a designer after manufacturing in a factory. An FPGA executes
any classical algorithms. On the other hand, our variational universal
quantum-state generator creates an arbitrary quantum state. We program by
using the variational parameters $\theta _{i}$. In this sense, the above
quantum circuit may be considered as a quantum generalization of FPGA, which
is a quantum FPGA (q-FPGA).

We show explicitly how the cost function is renewed for each variational
step in the case of two- and three-qubit universal quantum circuits in Fig.\ref{FigCost}, 
where we have generated the initial and the final states
randomly. We optimize 15 parameters $\theta _{i}$\ for two-qubit universal
quantum circuits and 82 parameters $\theta _{i}$\ for three-qubit universal
quantum circuits. We find that $U_{\text{ini-fin}}$ is well determined by
variational method as in Fig.\ref{FigCost}.

\medskip \noindent \textbf{Variational quantum support vector machine.}\smallskip

We demonstrate a binary classification problem in two dimensions based on
the support vector machine. We prepare a data set, where red points have a
distribution around $\left( r\cos \Theta ,r\sin \Theta \right) $ with
variance $r$, while blue points have a distribution around $\left( -r\cos
\Theta ,-r\sin \Theta \right) $ with variance $r$. We assume the Gaussian
normal distribution. We choose $\Theta $ randomly. We note that there are
some overlaps between the red and blue points, which is the soft margin
model.

As an example, we show the distribution of red and blue points and the lines
obtained by the variational method marked in cyan and by the direct solution
of (\ref{LinearEq}) marked in magenta in Fig.\ref{FigSVM}. They agrees well
with one another, where both of the lines well separate red and blue points.
We have prepared 31 red points and 32 blue points, and used six
qubits.\medskip

\medskip \noindent \textbf{\large Discussion}\medskip

We have proposed that the matrix $F$ is efficiently inputted into a quantum
computer by using the $\Gamma $-matrix expansion method. There are many ways
to use a matrix in a quantum computer such as linear regression and
principal component analysis. Our method will be applicable to these cases.

Although it is possible to obtain the exact solution for the linear equation
by the HHL algorithm, it requires many gates. On the other hand, it is often
hard to obtain the exact solution by variational methods since
trial functions may be trapped to a local minimum. However, this problem is
not serious for the machine learning problem because it is more important to
obtain an approximate solution efficiently rather than an exact solution by using many gates.
Indeed, our optimized hyperplane also well separates red and blue points as
shown in Fig.\ref{FigSVM}(a).

In order to classify $M$ data, we need to prepare $\log _{2}M$ qubits. It is
hard to execute a large number of data points by current quantum computers.
Recently, it is shown that electric circuits may simulate universal quantum
gates\cite{EzawaUniv,EzawaDirac,LCBit} based on the fact that the Kirchhoff
law is rewritten in the form of the Schr\"{o}dinger equation\cite{EzawaSch}.
Our variational algorithm will be simulated by using them.

\medskip\noindent\textbf{\large Methods}

\medskip \noindent \textbf{A: Support vector machine.}\smallskip

\begin{figure}[t]
\centerline{\includegraphics[width=0.88\textwidth]{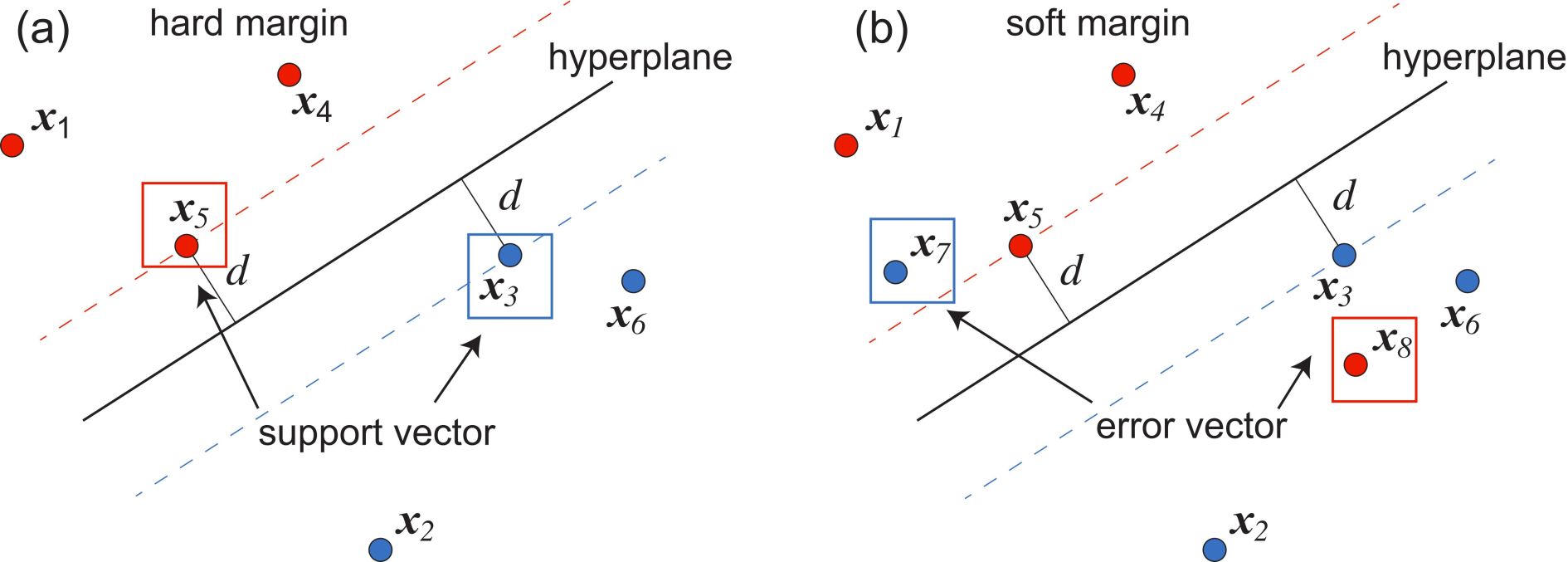}}
\caption{Illustration of the hyperplane and the support vector. Two support
vectors are marked by red and blue squares. (a) Hard margin where red and
blue points are separated perfectly,  and (b) soft margin where they are separated imperfectly.}
\label{FigSV}
\end{figure}

A support vector machine is an algorithm for supervised learning\cite{Vap,Noble,Suy}. 
We first prepare a set of training data, where each point
is marked either in red or blue. Then, we determine a hyperplane separating
red and blue points. After learning, input data are classified into red or
blue by comparing the input data with the hyperplane. The support vector
machine maximizes a margin, which is a distance between the hyperplane and
data points. If red and blue points are perfectly separated by the
hyperplane, it is called a hard margin problem [Fig.\ref{FigSV}(a)].
Otherwise, it is called a soft margin problem [Fig.\ref{FigSV}(b)].

We minimize the distance $d_{j}$\ between a data point $\boldsymbol{x}_{j}$ and
the hyperplane given by 
\begin{equation}
d_{j}=\frac{\left\vert \boldsymbol{\omega }\cdot \boldsymbol{x}_{j}+\omega
_{0}\right\vert }{\left\vert \boldsymbol{\omega }\right\vert }.
\end{equation}%
We define support vectors $\boldsymbol{x}$ as the closest points to the
hyperplane. There is such a vector in each side of the hyperplane, as shown
in Fig.\ref{FigSV}(a). This is the origin of the name of the support vector
machine. Without loss of generality, we set 
\begin{equation}
\left\vert \boldsymbol{\omega }\cdot \boldsymbol{x}+\omega _{0}\right\vert =1
\label{EqApA}
\end{equation}%
for the support vectors, because the hyperplane is present at the
equidistance of two closest data points and because it is possible to set
the magnitude of $\left\vert \boldsymbol{\omega }\cdot \boldsymbol{x}+\omega
_{0}\right\vert $\ to be $1$ by scaling $\boldsymbol{\omega }$ and $\omega _{0}$. 
Then, we maximize the distance 
\begin{equation}
d=\frac{\left\vert \boldsymbol{\omega }\cdot \boldsymbol{x}+\omega _{0}\right\vert }{%
\left\vert \boldsymbol{\omega }\right\vert }=\frac{1}{\left\vert \boldsymbol{\omega }%
\right\vert },
\end{equation}%
which is identical to minimize $\left\vert \boldsymbol{\omega }\right\vert $.

First, we consider the hard margin problem, where red and blue points are
perfectly separable. All red points satisfy $\boldsymbol{\omega }\cdot \boldsymbol{x}%
_{j}+\omega _{0}>1$ and all blue points satisfy $\boldsymbol{\omega }\cdot 
\boldsymbol{x}_{j}+\omega _{0}<-1$. We introduce variables $y_{j}$, 
where $y_{j}=1$ for red points and $y_{j}=-1$ for blue points. Using them, the
condition is rewritten as%
\begin{equation}
\left( \boldsymbol{\omega }\cdot \boldsymbol{x}_{j}+\omega _{0}\right) y_{j}\geq 1
\end{equation}%
for each $j$. The problem is reduced to find the minimum of $\left\vert 
\boldsymbol{\omega }\right\vert ^{2}$ under the above inequalities. The
optimization under inequality conditions is done by the Lagrange multiplier
method with the Karush-Kuhn-Tucker condition\cite{KKT}. It is expressed in
terms of the Lagrangian as%
\begin{equation}
L\left( \boldsymbol{\omega },\omega _{0},\boldsymbol{\alpha }\right) =\frac{1}{2}%
\left\vert \boldsymbol{\omega }\right\vert ^{2}-\sum_{j}\beta _{j}[\left( 
\boldsymbol{\omega }\cdot \boldsymbol{x}_{j}+\omega _{0}\right) y_{j}-1],
\end{equation}%
where $\beta _{j}$ are Lagrange multipliers to ensure the constraints.

For the soft margin case, we cannot separate two classes exactly. In order
to treat this case, we introduce slack variables $\xi _{j}$ satisfying%
\begin{equation}
\left( \boldsymbol{\omega }\cdot \boldsymbol{x}_{j}+\omega _{0}\right) y_{j}\geq
1-\xi _{j},\qquad \xi _{j}\geq 0
\end{equation}%
and redefine the cost function as%
\begin{equation}
E_{\text{cost}}=\frac{1}{2}\left\vert \boldsymbol{\omega }\right\vert
^{2}+\gamma \sum_{j=1}^{M}\xi _{j}^{2}.
\end{equation}%
Here, $\gamma =\infty $ corresponds to the hard margin. The second term
represents the penalty for some of data points to have crossed over the
hyperplane. The Lagrangian is modified as%
\begin{equation}
L\left( \boldsymbol{\omega },\omega _{0},\xi _{i},\boldsymbol{\beta }\right) 
=\frac{1}{2}\left\vert \boldsymbol{\omega }\right\vert ^{2}+\gamma \sum_{j=1}^{M}\xi
_{j}^{2}-\sum_{j=1}^{M}\left[ \left( \boldsymbol{\omega }\cdot \boldsymbol{x}_{j}
+\omega _{0}\right) \beta _{j}y_{j}-\left( 1-\xi _{i}\right) \right] .
\end{equation}%
The stationary points are determined by%
\begin{align}
\frac{\partial L}{\partial \boldsymbol{\omega }} =&\boldsymbol{\omega }%
-\sum_{j=1}^{M}\beta _{j}y_{j}\boldsymbol{x}_{j}=0,  \label{Lw} \\
\frac{\partial L}{\partial \omega _{0}} =&-\sum_{j=1}^{M}\beta _{j}y_{j}=0,
\\
\frac{\partial L}{\partial \xi _{j}} =&\gamma \xi _{j}-\beta _{j}=0,
\label{Lz} \\
\frac{\partial L}{\partial \beta _{j}} =&\left( \boldsymbol{\omega }\cdot 
\boldsymbol{x}_{j}+\omega _{0}\right) y_{j}-\left( 1-\xi _{i}\right) =0.
\label{Lb}
\end{align}%
We may solve these equations to determine $\boldsymbol{\omega }$ and $\nu _{j}$
as%
\begin{equation}
\boldsymbol{\omega }=\sum_{j=1}^{M}\beta _{j}y_{j}\boldsymbol{x}_{j},
\end{equation}%
from (\ref{Lw}), and 
\begin{equation}
\xi _{j}=\beta _{j}/\gamma
\end{equation}%
from (\ref{Lz}). Inserting them into (\ref{Lb}), we find%
\begin{equation}
y_{j}\sum_{i=1}^{M}\left( \beta _{i}y_{i}\boldsymbol{x}_{i}\cdot \boldsymbol{x}_{j}
+\omega _{0}\right) -\left( 1-\beta _{j}/\gamma \right) =0.
\end{equation}%
Since $y_{j}^{2}=1$, it is rewritten as%
\begin{equation}
\omega _{0}+\sum_{i=1}^{M}\left( \boldsymbol{x}_{i}\cdot \boldsymbol{x}_{j}+\delta
_{ij}/\gamma \right) \beta _{i}y_{i}=y_{j}.
\end{equation}%
Since $\beta _{j}$ appears always in a pair with $y_{j}$, we introduce a new
variable defined by 
\begin{equation}
\alpha _{j}=\beta _{j}y_{j},
\end{equation}%
and we define the Kernel matrix $K_{ij}$ as%
\begin{equation}
K_{ij}=\boldsymbol{x}_{i}\cdot \boldsymbol{x}_{j}.
\end{equation}%
Then, $\omega _{0}$ and $\alpha _{j}$ are obtained by solving linear
equations%
\begin{align}
\sum_{i=1}^{M}\alpha _{j} =&0, \\
\omega _{0}+\sum_{i=1}^{M}\left( \boldsymbol{x}_{i}\cdot \boldsymbol{x}_{j}+\delta
_{ij}/\gamma \right) \alpha _{i} =&y_{j},
\end{align}%
which are summarized as%
\begin{equation}
\left( 
\begin{array}{cccc}
0 & 1 & \cdots & 1 \\ 
1 &  &  &  \\ 
\vdots &  & K+I_{M}/\gamma &  \\ 
1 &  &  & 
\end{array}%
\right) \left( 
\begin{array}{c}
\omega _{0} \\ 
\alpha _{1} \\ 
\vdots \\ 
\alpha _{M}%
\end{array}%
\right) =\left( 
\begin{array}{c}
0 \\ 
y_{1} \\ 
\vdots \\ 
y_{M}%
\end{array}%
\right) ,
\end{equation}%
which is Eq.(\ref{LinearEq}) in the main text. Finally, $\boldsymbol{\omega }$
is determined by%
\begin{equation}
\boldsymbol{\omega }=\sum_{j=1}^{M}\alpha _{j}\boldsymbol{x}_{j}.
\end{equation}%
Once the hyperplane is determined, we can classify new input data into red if%
\begin{equation}
\boldsymbol{\omega }\cdot \boldsymbol{x}_{j}+\omega _{0}>0
\end{equation}%
and blue if 
\begin{equation}
\boldsymbol{\omega }\cdot \boldsymbol{x}_{j}+\omega _{0}<0.
\end{equation}%
Thus, we obtain the hyperplane for binary classification.

\medskip \noindent \textbf{B: $\Gamma $ matrix expansion.}\smallskip

We explicitly show how to calculate $c_{j}$ in (\ref{cj}) based on the $\Gamma $ matrix expansion for the one and two qubits.

\smallskip\textbf{One qubit:}

We show an explicit example of the $\Gamma $-matrix expansion for one qubit.
Ome qubit is represented by a $2\times 2$ matrix,%
\begin{equation}
F=\left( 
\begin{array}{cc}
F_{11} & F_{12} \\ 
F_{21} & F_{22}%
\end{array}%
\right) .
\end{equation}%
The column vectors are explicitly given by%
\begin{align}
\left\vert f_{1}\right\rangle =&\left( 
\begin{array}{c}
F_{11} \\ 
F_{21}%
\end{array}%
\right) =F_{11}\left\vert 0\right\rangle +F_{21}\left\vert 1\right\rangle
,\qquad \\
\left\vert f_{2}\right\rangle =&\left( 
\begin{array}{c}
F_{12} \\ 
F_{22}%
\end{array}%
\right) =F_{12}\left\vert 0\right\rangle +F_{22}\left\vert 1\right\rangle .
\end{align}%
The coefficient $c_{j}$ in (\ref{cj}) is calculated as%
\begin{equation}
c_{j}=\text{Tr}\left[ \sigma _{j}F\right] =\langle 0|\sigma _{j}\left\vert
f_{1}\right\rangle +\left\langle 1\right\vert \sigma _{j}\left\vert
f_{2}\right\rangle =\sum_{p=0,1}\langle p|\sigma _{j}\left\vert
f_{p}\right\rangle =\sum_{p=0,1}\langle 0|U_{X}^{\left( p\right) }\sigma
_{j}U_{f_{p}}\left\vert 0\right\rangle .
\end{equation}

\smallskip\textbf{Two qubits:}

Next, we show an explicit example of the $\Gamma $-matrix expansion for two
qubits. Two qubits are represented by a $4\times 4$ matrix,

\begin{equation}
F=\left( 
\begin{array}{cccc}
F_{11} & F_{12} & F_{13} & F_{14} \\ 
F_{21} & F_{22} & F_{23} & F_{24} \\ 
F_{31} & F_{32} & F_{33} & F_{34} \\ 
F_{41} & F_{42} & F_{43} & F_{44}%
\end{array}%
\right) .
\end{equation}%
The column vectors are explicitly given by%
\begin{align}
\left\vert f_{1}\right\rangle  =&\left( 
\begin{array}{c}
F_{11} \\ 
F_{21} \\ 
F_{31} \\ 
F_{41}%
\end{array}%
\right) =F_{11}\left\vert 00\right\rangle +F_{21}\left\vert 01\right\rangle
+F_{31}\left\vert 10\right\rangle +F_{41}\left\vert 11\right\rangle , \\
\left\vert f_{2}\right\rangle  =&\left( 
\begin{array}{c}
F_{12} \\ 
F_{22} \\ 
F_{32} \\ 
F_{42}%
\end{array}%
\right) =F_{12}\left\vert 00\right\rangle +F_{22}\left\vert 01\right\rangle
+F_{32}\left\vert 10\right\rangle +F_{42}\left\vert 11\right\rangle , \\
\left\vert f_{3}\right\rangle  =&\left( 
\begin{array}{c}
F_{13} \\ 
F_{23} \\ 
F_{33} \\ 
F_{43}%
\end{array}%
\right) =F_{13}\left\vert 00\right\rangle +F_{23}\left\vert 01\right\rangle
+F_{33}\left\vert 10\right\rangle +F_{43}\left\vert 11\right\rangle , \\
\left\vert f_{4}\right\rangle  =&\left( 
\begin{array}{c}
F_{14} \\ 
F_{24} \\ 
F_{34} \\ 
F_{44}%
\end{array}%
\right) =F_{14}\left\vert 00\right\rangle +F_{24}\left\vert 01\right\rangle
+F_{34}\left\vert 10\right\rangle +F_{44}\left\vert 11\right\rangle .
\end{align}%
The coefficient $c_{j}$ in (\ref{cj}) is calculated as%
\begin{equation}
c_{j}=\text{Tr}\left[ \Gamma _{j}F\right] =\left\langle 00\right\vert \Gamma
_{j}\left\vert f_{1}\right\rangle +\left\langle 01\right\vert \Gamma
_{j}\left\vert f_{2}\right\rangle +\left\langle 10\right\vert \Gamma
_{j}\left\vert f_{3}\right\rangle +\left\langle 11\right\vert \Gamma
_{j}\left\vert f_{4}\right\rangle =\sum_{p=0}^{3}\langle \!\left\langle
p\right\vert \Gamma _{j}\left\vert f_{p}\right\rangle =\sum_{p=0}^{3}\langle
\!\left\langle 0\right\vert U_{X}^{\left( p\right) }\Gamma
_{j}U_{f_{p}}\left\vert 0\right\rangle \!\rangle .
\end{equation}

\medskip \noindent \textbf{C: Universal quantum circuits.}\smallskip

Angle variables are used as variational parameters in a universal quantum
circuit learning. We present examples for one, two and three qubits.

\smallskip\textbf{One-qubit universal quantum circuit: }

\begin{figure*}[t]
\centerline{\includegraphics[width=0.98\textwidth]{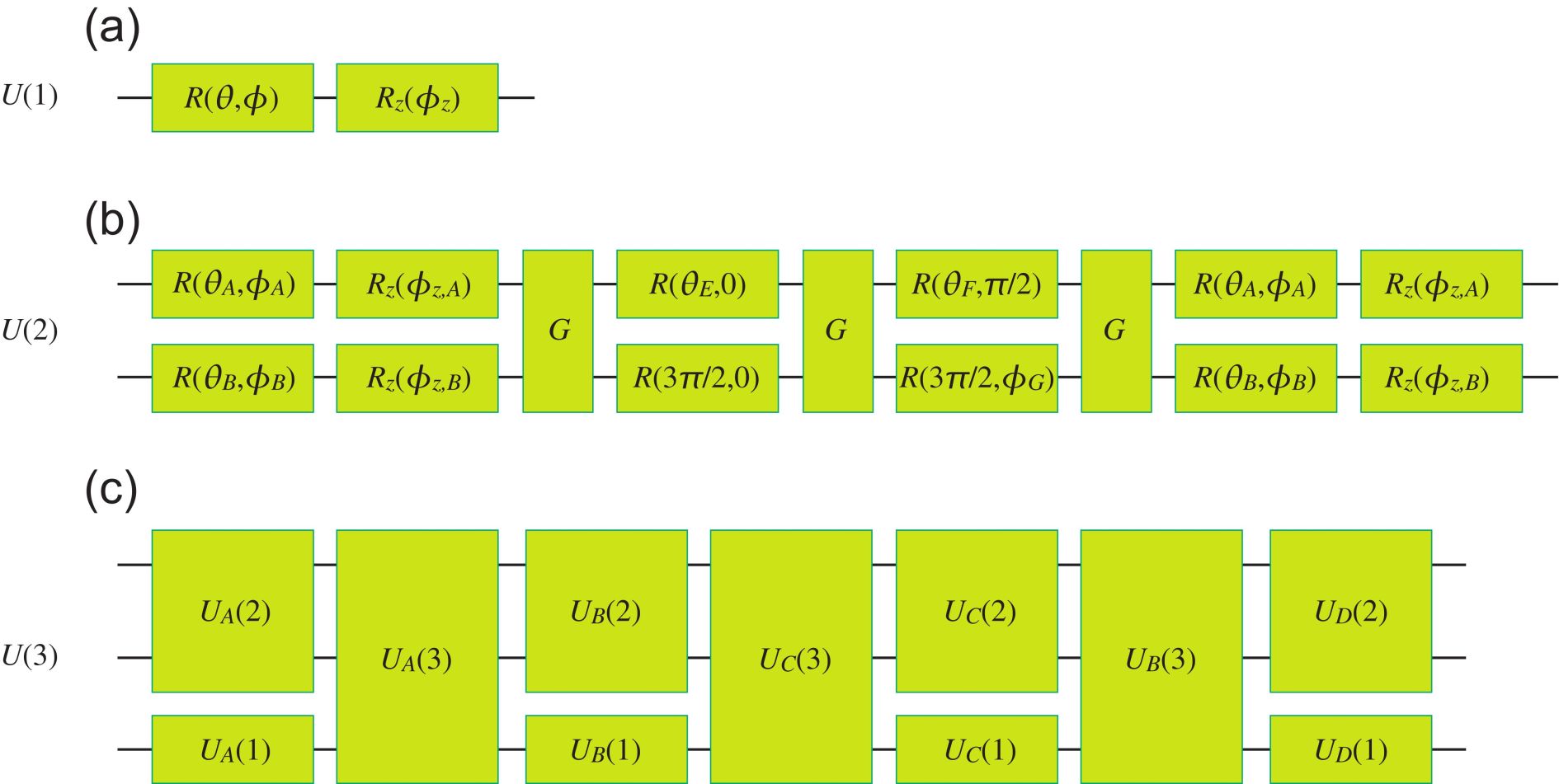}}
\caption{{}Universal quantum circuits for (a) one, (b) two and (c) three
qubits.}
\label{FigUniv}
\end{figure*}

The single-qubit rotation gates are defined by%
\begin{align}
R\left( \theta ,\phi \right)  =&\exp \left[ -i\theta \left( \sigma _{x}\cos
\phi +\sigma _{y}\sin \phi \right) /2\right] , \\
R_{z}\left( \phi _{z}\right)  =&\exp \left[ -i\sigma _{z}\phi _{z}/2\right].
\end{align}%
The one-qubit universal quantum circuit is constructed as%
\begin{equation}
U^{\left( 1\right) }\left( \theta ,\phi ,\phi _{z}\right) =R\left( \theta
,\phi \right) R_{z}\left( \phi _{z}\right) =\left( 
\begin{array}{cc}
e^{-i\phi _{z}/2}\cos \frac{\theta }{2} & -ie^{i\left( \phi _{z}/2-\phi
\right) }\sin \frac{\theta }{2} \\ 
-ie^{-i\left( \phi _{z}/2-\phi \right) }\sin \frac{\theta }{2} & e^{i\phi
_{z}/2}\cos \frac{\theta }{2}%
\end{array}%
\right) .
\end{equation}%
We show a quantum circuit in Fig.\ref{FigUniv}(a). There are three
variational parameters.

It is obvious that an arbitrary state is realized starting from the state $%
\left\vert 0\right\rangle $ as%
\begin{equation}
U\left( 1\right) \left( 
\begin{array}{c}
1 \\ 
0%
\end{array}%
\right) =\left( 
\begin{array}{c}
e^{-i\phi _{z}/2}\cos \frac{\theta }{2} \\ 
-ie^{-i\left( \phi _{z}/2-\phi \right) }\sin \frac{\theta }{2}%
\end{array}%
\right) .
\end{equation}

\smallskip\textbf{Two-qubit universal quantum circuit:}

The two-qubit universal quantum circuit is constructed as\cite{Hanne}%
\begin{align}
U\left( 2\right) \equiv &\left[ U^{\left( 1\right) }\left( \theta _{A},\phi
_{A},\phi _{z,A}\right) \otimes U^{\left( 1\right) }\left( \theta _{B},\phi
_{B},\phi _{z,B}\right) \right] U_{G}\left[ R\left( \theta _{E},0\right)
\otimes R\left( \frac{3\pi }{2},0\right) \right] U_{G}\left[ R\left( \theta
_{F},\frac{\pi }{2}\right) \otimes R\left( \frac{3\pi }{2},\theta
_{G}\right) \right]  \notag \\
&U_{G}\left[ U^{\left( 1\right) }\left( \theta _{C},\phi _{C},\phi
_{z,C}\right) \otimes U^{\left( 1\right) }\left( \theta _{D},\phi _{D},\phi
_{z,D}\right) \right] ,
\end{align}%
where the entangling two-qubit gate is defined by\cite{Hanne}%
\begin{equation}
U_{G}=e^{-i\pi /4}\exp \left[ \frac{i\pi }{4}\sigma _{z}\otimes \sigma _{z}\right] .
\end{equation}%
The two-qubits universal quantum circuit contains 15 variational parameters.
We show a quantum circuit in Fig.\ref{FigUniv}(b).

\smallskip\textbf{Three-qubit universal quantum circuit:}

The three-qubit universal quantum circuit is constructed as%
\begin{equation}
U\left( 3\right) \equiv \left[ U_{A}^{\left( 2\right) }\otimes U_{A}^{\left(
1\right) }\right] U_{A}\left( 3\right) \left[ U_{B}^{\left( 2\right)
}\otimes U_{B}^{\left( 1\right) }\right] U_{C}\left( 3\right) \left[
U_{C}^{\left( 2\right) }\otimes U_{C}^{\left( 1\right) }\right] U_{B}\left(
3\right) \left[ U_{D}^{\left( 2\right) }\otimes U_{D}^{\left( 1\right) }%
\right] ,
\end{equation}%
where $U_{A}^{\left( 1\right) }$, $U_{B}^{\left( 1\right) }$, $U_{C}^{\left(
1\right) }$, and $U_{D}^{\left( 1\right) }$\ are one-qubit universal quantum
circuits, while $U_{A}^{\left( 2\right) }$, $U_{B}^{\left( 2\right) }$, 
$U_{C}^{\left( 2\right) }$, and $U_{D}^{\left( 2\right) }$\ are two-qubit
universal quantum circuit and 
\begin{align}
U_{A}\left( 3\right)  =&\exp \left[ i\left( \theta _{xxz}^{A}\sigma
_{x}\otimes \sigma _{x}\otimes \sigma _{z}+\theta _{yyz}^{A}\sigma
_{x}\otimes \sigma _{x}\otimes \sigma _{z}+\theta _{zzz}^{A}\sigma
_{z}\otimes \sigma _{z}\otimes \sigma _{z}\right) \right] , \\
U_{B}\left( 3\right)  =&\exp \left[ i\left( \theta _{xxz}^{B}\sigma
_{x}\otimes \sigma _{x}\otimes \sigma _{z}+\theta _{yyz}^{B}\sigma
_{x}\otimes \sigma _{x}\otimes \sigma _{z}+\theta _{zzz}^{B}\sigma
_{z}\otimes \sigma _{z}\otimes \sigma _{z}\right) \right] , \\
U_{C}\left( 3\right)  =&\exp \left[ i\left( \theta _{xxx}^{C}\sigma
_{x}\otimes \sigma _{x}\otimes \sigma _{x}+\theta _{yyx}^{C}\sigma
_{y}\otimes \sigma _{y}\otimes \sigma _{x}+\theta _{zzx}^{C}\sigma
_{z}\otimes \sigma _{z}\otimes \sigma _{x}+\theta _{00x}^{C}\sigma
_{0}\otimes \sigma _{0}\otimes \sigma _{x}\right) \right] .
\end{align}%
Eplicit quantum circuits for $U_{A}\left( 3\right) $, $U_{B}\left( 3\right) $
and $U_{C}\left( 3\right) $\ are shown in Ref.\cite{Sousa}. The three-qubits
universal quantum circuit contains 82 variational parameters. We show a
quantum circuit in Fig.\ref{FigUniv}(c).

\smallskip\textbf{Multi-qubit universal quantum circuit:}

General multi-qubit universal quantum circuit is constructed in Ref.\cite{Motto}. 
The minimum numbers of variational parameters are $4^{N}-1$\ for 
$N$-qubit unicersal quantum circuits. However, we need more variational
parameters in the currently known algorithm for $N\geq 3$.

\newpage
\medskip \noindent \textbf{\large Acknowledgements}\medskip

The author is very much grateful to E. Saito and N. Nagaosa for helpful discussions on
the subject. This work is supported by the Grants-in-Aid for Scientific
Research from MEXT KAKENHI (Grants No. JP17K05490 and No. JP18H03676). This
work is also supported by CREST, JST (JPMJCR16F1 and JPMJCR20T2).

\medskip \noindent \textbf{\large Author contributions}\medskip

M.E. conceived the idea, performed the analysis, and wrote the manuscript.

\medskip \noindent \textbf{\large Additional information}\medskip

\textbf{Competing financial and non-financial interests:} The author declares no competing financial and
non-financial interests.
\end{document}